\documentclass[prl, reprint, superscriptaddress]{revtex4-2}

\usepackage{graphicx}
\usepackage{dcolumn}
\usepackage{bm}
\usepackage{xcolor}
\begin{document}


\title{Flat Band and Many-body Gap in Chirally Twisted Triple Bilayer Graphene}

\author{Wenlu Lin}
\thanks{These authors contributed equally to this study.}
\affiliation{International Center for Quantum Materials, 
  Peking University, Haidian, Beijing 100871, China}
\author{Wenxuan Wang}
\thanks{These authors contributed equally to this study.}
\affiliation{International Center for Quantum Materials, 
  Peking University, Haidian, Beijing 100871, China}
\author{Shimin Cao}
\thanks{These authors contributed equally to this study.}
\affiliation{International Center for Quantum Materials, 
  Peking University, Haidian, Beijing 100871, China}
\author{Miao Liang} 
\affiliation{School of Physics and Wuhan National High Magnetic Field Center,
Huazhong University of Science and Technology, Wuhan 430074, China}
\author{Lili Zhao} 
\affiliation{International Center for Quantum Materials, 
Peking University, Haidian, Beijing 100871, China} 
\author{Kenji Watanabe} 
\affiliation{Research Center for Electronic and Optical Materials, National Institute of Material Sciences, 1-1 Namiki, Tsukuba 305-0044, Japan} 
\author{Takashi Taniguchi} 
\affiliation{Research Center for Materials Nanoarchitectonics, National Institute of Material Sciences, 1-1 Namiki, Tsukuba 305-0044, Japan}

\author{Jinhua Gao} 
\affiliation{School of Physics and Wuhan National High Magnetic Field Center,
Huazhong University of Science and Technology, Wuhan 430074, China}
\affiliation{Hefei National Laboratory, Hefei 230088, China}
\author{Jianhao Chen} 
\email{chenjianhao@pku.edu.cn} 
\affiliation{International Center for Quantum Materials, 
  Peking University, Haidian, Beijing 100871, China} 
\author{Xiaobo Lu} 
\email{xiaobolu@pku.edu.cn} 
\affiliation{International Center for Quantum Materials, 
  Peking University, Haidian, Beijing 100871, China} 
\author{Yang Liu} 
\email{liuyang02@pku.edu.cn} 
\affiliation{International Center for Quantum Materials, 
  Peking University, Haidian, Beijing 100871, China}
\affiliation{Hefei National Laboratory, Hefei 230088, China}

\date{\today}

\begin{abstract}

  We experimentally investigate the band structures of chirally
  twisted triple bilayer graphene. The new kind of moir\'{e}
  structure, formed by three pieces of helically stacked Bernal
  bilayer graphene, has flat bands at charge neutral point based on
  the continuum approximation. We experimentally confirm the existence
  of flat bands and directly acquire the gap in-between flat bands as
  well as between the flat bands and dispersive bands from the
  capacitance measurements. We discover a finite gap even at zero
  perpendicular electric field, possibly induced by the Coulomb
  interaction and ferromagnetism. Our quantitative study not only
  provides solid evidence for the flat-band and interesting physics,
  but also introduces a quantitative approach to explore phenomena of
  similar moir\'{e} systems.

\end{abstract}

\maketitle

The relation between the chemical potential $\mu$ and the particle
density $n$ is one of the essential properties describing a Fermionic
system \cite{Eisenstein.PRB.1994, Hajaj.PRB.2013}. Many interesting
physics is related to anomalies in this $\mu$ vs. $n$ relation, such
as the van-Hove singularity and Dirac cone where $dn/d\mu$ divergies
or vanishes \cite{VH.PRL.2010, Ashoori.PRL.2019}. External magnetic
field, artificial patterns or electron-phonon interactions can also
induce similar anomalies such as heavy Fermions, superconductivity,
moir\'{e} supperlattice, etc. \cite{LGH.NP.2010,
Tomarken.PRL.2019}. Many probing methods, e.g. conductivity and
commensurability oscillation, etc. \cite{CO.EuLetters.1989,
CO.PRL.2018}, can reveal features with proper assumptions such as
uniform and constant scattering rate. However, a direct probe of these
features is always of fundamental importance for exploring condensed matter
systems.

Recently, interests increasingly grow in studying twistronics
materials, where the modulated interlayer coupling leads to abundant
features, including flat-band and many-body gaps and so on
\cite{Allan.PNAS.2011, Yacoby.NP.2021}. The appearance of flat-band is
generally believed to be the onset of many interaction induced quantum
phases such as fractional quantum Hall effect and Wigner crystal
\cite{PerspectiveQHE.1998}. Signatures of superconductivity-like
phenomenon and charge-density waves (so-called ``generalized'' Wigner
crystal) have been reported in these systems \cite{WWX.PRL.2024}. The
chirally twisted triple bilayer graphene (CTTBG) has two equal-sized
moir\'{e} Brillouin zones with a mis-orientation of the rotation
angle, generated by the two sets of moir\'{e} superlattice
\cite{CTTBG.SA.2023, CTTBG.arXiv.2023}. The ultra-flat moir\'{e} bands
can exist for a relatively wide range of the rotation angle, and are
well isolated from other dispersive bands at higher energy, making
CTTBG a new platform for exploring interacting physics. When a finite
perpendicular electric field is applied, the intertwined flat valance
and conduction bands separate and a band gap at charge neutral point
(CNP) develops.

Despite the criticality of the flat-bands in twistronics devices, firm
and direct experimental demonstration of its existence and a direct
measure of the moir\'{e} supperlattice strength are still missing. In
this work, we compare the capacitance measured from` monolayer
graphene (MLG) and CTTBG. We evidence that CTTBG has flat bands
coexisting with many-body ferromagnetism gap whose value we can
accurately acquire. This intrinsic band gap has been reported in other
suspended graphene systems while has not been discovered in such
moir\'{e} systems \cite{Yacoby.Science.2010, LCN.NanoLetter.2012}.

 \begin{figure}
 \includegraphics[scale=1]{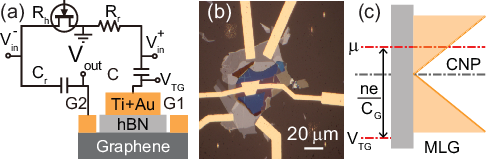}%
 \caption{\label{fig0} (a) Schematic diagram of our setup, which measures the capacitance between the top gate (TG) and the contacts. (b) Picture of our monolayer graphene (MLG) sample. (c) Cartoon explaining the measurement principle. The orange shades represent the density of state (DOS) of MLG.}
\end{figure}

Our CTTBG consists of three pieces of Bernal stacked bilayer graphene
chirally stacked on top of each other with the same rotation
angle. The two twisted angles are $1.7^{\circ}$ so that the
supermoir\'{e} structure with ultra-large periodicity largely retains
the electronic properties dominated by original lattice. The CTTBG is
predicted to host a pair of flat moir\'{e} bands well isolated from
other dispersive bands at higher energy \cite{LM.PRB.2022,
WWX.PRL.2024}, giving rise to fundamentally different moir\'{e} band
structures and leading to exotic interaction-induced quantum
phenomena.

The measurements are performed in a dilution refrigerator whose base
temperature is below 10 mK. The longitudinal resistance $R_{xx}$ shown
in SI is measured using a standard lock-in technique ($<$ 30 Hz)
\cite{SM}. The capacitance measurements are conducted between top gate
(TG) and contacts using a cryogenic bridge, see Fig. 1(a). Capacitance
and conductance components can be simultaneously extracted in our
capacitance measurements. We have subtracted the parasitic capacitance
$C_{p}$ (typically about 30 fF) from all present data in this
manuscript as long as it's possible and necessary. We measure the
parasitic capacitance $C_p$ by tuning the graphene density to nearly
zero and applying a large perpendicular magnetic field (the inset of
Fig. 2(b)) when the graphene layer becomes insulating and does not
contribute to the capacitance. The Fig. 3(c-d) and 4(c-d) data is
taken at constant $D=0$ while the Fig. 4(a) data is taken by sweeping
$V_{FG}$ with constant $V_{BG}$. In these two cases, the gap $\Delta$
can be directly read from the width of this minimum $\Delta V_{TG}$
through $\Delta = \frac{2K}{1+K}\times \Delta V_{TG}$ or $\Delta =
\frac{K}{1+K}\times \Delta V_{TG}$, respectively.

We first demonstrate our measurement principle using a MLG device. The
TG is separated from the graphene by a thin h-BN layer \footnote{For
simplicity, we always define the gate bias $V_{G}$ and the graphene
chemical potential $\mu$ in reference to the condition when the
graphene is tuned to its CNP throughout this manuscript, see the black
dash-dotted line in Fig.  1(c).}. The accumulated charge $ne$ in the
two layers ($e$ is the electron charge and $n$ is the carrier density)
supports an electric potential difference across the insulating layer
through the geometric capacitance $C_G$, as well as changes the
graphene chemical potential by $\mu$, see Fig. 1(c). The voltage
between the graphene and the gate is the sum of these two
components. If the device has an additional bottom gate (BG), the
voltage-density relation can be expressed as: \begin{equation}
\left(\mathrm{V}_{\mathrm{TG}}-\mu\right) \times C_G+\left(V_{B
G}-\mu\right) \times C_G \times K^{-1}=n e \end{equation} where
$C_{G}$ is the geometric capacitance (measured between TG and the
graphene), $V_{TG}$ and $V_{BG}$ are the voltage between the
corresponding gates and the graphene. $K$ is the ratio between the
gating efficiencies of the TG and BG, and $K = \infty$ if the sample
has no (or floating) BG.

\begin{figure}
\includegraphics[scale=1]{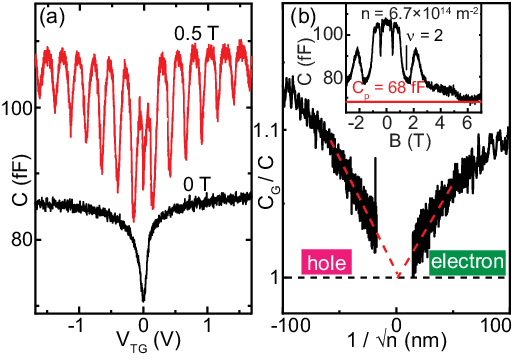}%
\caption{\label{fig1} (a) Capacitance data taken from MLG sample at B = 0 and 0.5 T. (b) The $C_{G} / C$ vs. $1/ \sqrt{n}$ plot of the zero magnetic field data in panel (a). The red dashed line is its linear fitting. The inset shows the $C$ vs. $B$ at density $n$ = $6.7 \times 10^{14} m^{-2}$. The parasitic capacitance $C_{p} \approx 68 fF$ is deduced from the minimum $C$ value at $B >$ 6 T.}
\end{figure}

\begin{figure*}
 \includegraphics[scale=1]{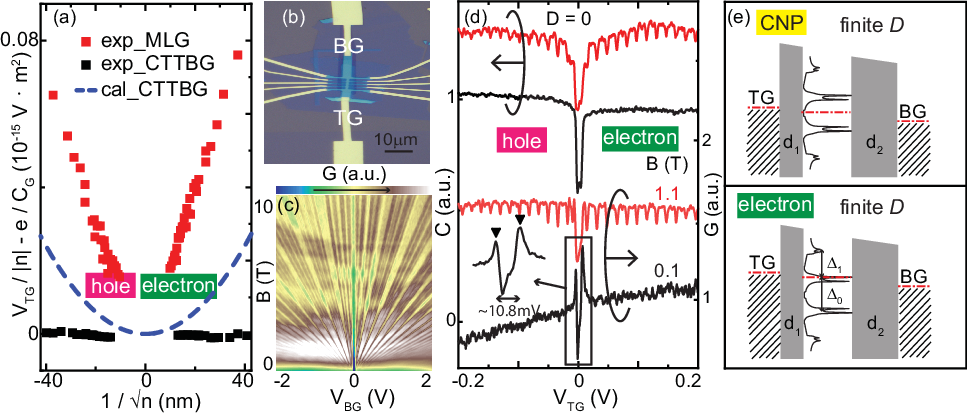}
 \caption{\label{fig2} (a) The $V_{TG} / n$ vs. $1/\sqrt{n}$ plot of the data from the MLG (red) and chirally twisted triple bilayer graphene (CTTBG, black) samples. The blue dashed line is the expected curve if the CTTBG sample has a single particle gap. (b) Picture of our CTTBG sample, which has local TG and global BG. (c) Landau fan of $G$ data at zero perpendicular electric field. (d) The $C$ and $G$ components vs. the TG voltage at different $B$. The BG voltage is swept simultaneously to keep the perpendicular electric field at zero. The inset shows the expanded $G$ near CNP where the triangle markers label two turning points. (e) Cartoons explaining our gap measurements. The red dotted lines represent the Fermi energy in graphene and the two gates. The DOS used in the cartoon is calculated for $D$ = 0.047 V/nm.. The flat band condition is defined as $V_{TG}$ = $V_{BG}$ = 0.}
\end{figure*}

The AC capacitance $C$ in Fig 2(a), measured by our bridge using a
small excitation \cite{ZLL.RSI.2022}, is the differencial capacitance
$C=e\frac{\partial n}{\partial V_{FG}}$. Near the Dirac cone, the
reduced compressibility $dn/d\mu$, i.e. the density of state (DOS) at
the Fermi energy, suppresses the graphene's charging capability and
leads to a capacitance minima in the zero-field trace of Fig. 2(a)
\cite{Hazeghi.RSI.2011, Zibrov.PRL.2018, Tomarken.PRL.2019,
AndreaY.PRL.2024}. Thus, the dispersion parameter can be derived from
the $dn/d\mu$ quantitatively if the capacitance can be measured with
high precision. We note that $C_G/C=1+(dn/d\mu)^{-1} \times
\frac{C_G}{e}$, and $dn/d\mu=\frac{2 e}{\hbar v_{\mathrm{F}}
\sqrt{\pi}} \sqrt{n}$ for linear dispersion Dirac Fermions with its
Fermi velocity $v_F$. Therefore, $v_F$ can be obtained from our
experimental data in the following fashion. We first plot the ratio
$C_G/C$ as a function of $1/\sqrt{|n|}$, the positive (negative) value
of the $1/\sqrt{|n|}$ corresponds to electron (hole). The results
exhibit linear dependence, directly evidencing the linear dispersion
of the Dirac Fermions. The fitting parameter $C_G$ is the device
geometric capacitance. It equals the $C$ measured at infinitely large
$n$ (and hence infinitely large $dn/d\mu$). Correct $C_G$ value sets
the y-axis intercept of the Fig. 2(b) data to unity. Finally, we can
calculate the Fermi velocity $v_{F}$ from the slope of the linear
fitting, see the red dashed lines in Fig. 2(b). $v_{F}$ equals $1.23
\times 10^{6} m/s$ for holes and $1.14 \times 10^{6} m/s$ for
electrons. It is worth emphasizing that, the density $n$ is not
proportional to the applied gate voltage because of the varying
$dn/d\mu$. Instead, the density used in Fig. 2(b) is measured
experimentally from quantum oscillations, which we will discuss later.

As pointed out in an earlier work, local compressibility measurement
is usually more sensitive than transport measurement
\cite{Smet.Yacoby.PRL.2013}. The capacitance data exhibits minimum as
soon as a small area of incompressible plaques form inside the
samples, while the usual transport measurement can only see features
when these plaques form a connected path. We are able to observe clear
quantum oscillations at very small magnetic field $\sim 0.1$ T. When
an interger number of Landau levels are occupied, the system's
compressibility reduces and a minima appears in the measured $C$
\footnote{The $C$ and $G$ are highly entangled with each other
\cite{ZLL.CPL.2022}. Fortunately, this effect does not change our
conclusion.}. We then use the filling factor $\nu$ and magnetic fields
$B$ of these minima to calculate the exact particle density through $n
= \frac{eB\nu}{h}$ without any fitting parameters. As shown in Eq. 1,
the ratio $V_{TG}/n$ approaches $eC_G$ as $|n|$ increases since
$\mu/n$ vanishes. We summarize the data read from oscillations at
fields ranging from 0.5 to 2 T in Fig. 3(a), and calculates $v_{F}$
from the slope of the linear relation $\mu/n \propto
1/\sqrt{|n|}$. $v_{F}$ is $0.89 \times 10^{6} m/s$ for holes and $1.04
\times 10^{6} m/s$ for electrons. In principle, our analysis in
Fig. 3(a) has no fitting parameters. It perfectly agrees with earlier
STM studies, except that we can now have the graphene encapsulated
between two gates \cite{ZSY.NP.2006, LGH.NP.2007, LGH.PRL.2009,
HL.PRB.2015}. This experimentally measured $n$ vs. gate voltage
relation is used in analyzing Fig. 2(b).

\begin{figure*}
 \includegraphics[scale=1]{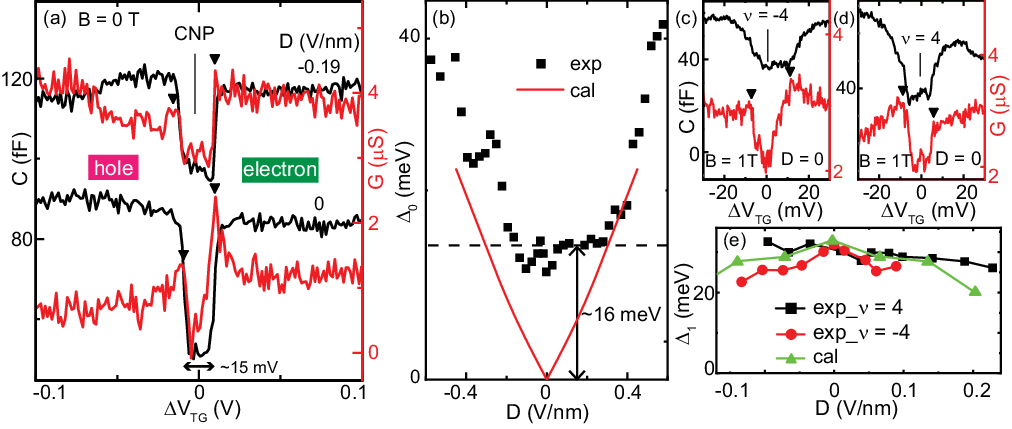}%
 \caption{\label{fig3} (a)(c-d) The $C$ and $G$ components from the CTTBG sample near CNP at $\nu = \pm 4$ and different $D$ and $B$ as labeled. Black triangles mark the turning points from which we determine the gaps. (b) The gaps at CNP extracted with different $D$. The red line shows the predicted gap by the continuum approximation. (e) Measured gaps at $\nu = \pm 4$ and their predicted values at different $D$.}
\end{figure*}

Once demonstrated the precision and reliability of this procedure, we
now perform similar measurements on our CTTBG sample, see
Fig. 3(a). The CTTBG sample has a local TG and a global BG, shown in
Fig. 3(b). Using these two gates, we can independently manipulate the
chemical potential $\mu$ (and hence the carrier density $n$) and the
out-of-plane electric field, which is quantified as electric
displacement $D$ (V/nm). In Fig. 3(c-d), we show the $G$ and $C$ by
sweeping the sample density while keeping zero $D$. Thanks to both the
high quality of our sample and high sensitivity of the measurement, we
are able to observe minima in both the $C$ and $G$ corresponding to
integer Landau level filling factors at magnetic field as small as 0.5
T. Therefore, following the same procedure of MLG, we summarize the
$V_G/n$ using black symbols in Fig. 3(a). Unlike the linear dependence
of $V_G/n$ on $1/\sqrt{n}$ seen in the MLG, we observe a constant
$V_G/n$ that is independent on the density, suggesting a constant
$\mu/n$ of either a parabolic dispersion or a diverging $dn/d\mu$ for
flat bands.

A gap-like feature, namely a minima in both the capacitance and
conductance, appears at the CNP, see Fig. 3(d). The conductance
minimum is flanked by two peaks, whose position coincides with the
abrupt change of $C$. The gap size $\Delta$ can be deduced from the
gate voltage difference between the two $G$ peaks as $\Delta =
\frac{2K}{1+K}\times \Delta V_{TG}$. If $\Delta$ is a single particle
gap, the experimental data points are expected to match the blue
dashed line. However, it is clearly not. Therefore, the gap seen in
the experiment is likely a manifestation of many-body ferromagnetism,
a signature that the interaction dominates over the kinetic
energy. Therefore, the non-single-particle gap seen at CNP as well as
the constant $V_G/|n|$ strongly evidence the existence of a group of
flat bands with different flavors \footnote{The collective pinning of
electron Wigner crystal may also lead to a gap-feature, however, it is
rather courageous to suggest this senario in our sample. Also, the
localization gap can also be excluded because our transport
measurement shows no insulating behavior; see Fig. S5.}.

The electric displacement $D$ breaks the inversion symmetry of the
system and induces a split $\Delta_0$ between the flat bands with
different valleys, see Fig. 3(e). The measurement principle can be
explained by Fig. 3(e), which shows two configurations when the
chemical potential $\mu$ lies at the CNP and in the flat band,
respectively \footnote{In Fig. 3(e), the electric field is kept to be constant, the gap size $\Delta$ is tuned by two electrical gates simultaneously, so that the ralation between $\Delta$ and $V_{TG}$ is $\Delta = \frac{2K}{1+K} \times V_{TG}$ while if the gap size is only tuned by $V_{TG}$, the ralation becomes $\Delta = \frac{K}{1+K} \times V_{TG}$.}. Opposite gate voltage bias $V_{TG}=-K^{-1}V_{BG}$
induces a finite $D$ while keeping $\mu$ unchanged, and non-zero
$V_{TG}+K^{-1}V_{BG}$ changes $\mu$. For example, Fig. 4(a) shows the $C$ and
$G$ minima seen at the CNP when $D$ = 0 and -0.19 V/nm. We summarize
the measured $\Delta_0$ as a function of $D$ in Fig. 4(b). The
experimental data points match the theoretical simulation (the
solid line) at large $D$. However, a clear finite gap is seen at
$D=0$. As discussed earlier, the gap at $D=0$ is likely the
spontaneous ferromagnetism stabilized by the dominating Coulomb
interaction. The $G$ minimum exists and its width
remains constant when a large magnetic field up to 10 T is applied,
see in Fig. 3(c). It is worth mentioning that we
observe no gap-feature at the CNP near $D=0$ in the transport study of
this sample \cite{WWX.PRL.2024}. It is likely that because the gap seen in capacitance measurements are charging
gaps, i.e. the discontinuity $\mu$ as a function of $n$. Therefore, it is not in conflict with the existence of percolation process in domain structures \cite{DePoortere.PRL.2003}.

The moir\'{e} flat bands are isolated from other dispersive bands by
the superlattice strength. The moir\'{e} gaps $\Delta_1$ in Fig. 3(e),
are another important parameter of twistronics materials, which is the
separation between flat bands from higher energy dispersing
bands. $\Delta_1$ gradually decreases as electric field increases
because electric field opens up the flat bands leading to the merging
of flat bands and dispersive bands. $\Delta_1$ corresponds to the
$\nu=\pm 4$ minimum in the $C$ and $G$ vs. $V_G$ trace, inside which
$n$ remains constant and $\mu$ varies through the gap, see
Fig. 4(c-d). We mark the turning points flanking the $G$ minima with
black triangles, which is used as a measure of the minimum
width. Fig. 4(e) summarizes the measured $\Delta_1$ as a function of
the $D$ at $\nu=\pm 4$. $\Delta_1$ gradually disappears as $D$
increases. The experimentally measured results are in good agreement
with the theoretical values, which validates our theoretical
parameters used in this study.

In conclusion, capacitance measurement as a local compressibility
probe has great advantage of revealing fine structure of
two-dimensional systems. The deduced Fermi velocity of MLG is in
agreement with previous results. Moreover, we study quantitative
intrinsic band structure of CTTBG. The inter-flat-band gap $\Delta_0$
exists at zero perpendicular electric field, possibly induced by
spontaneous ferromagnetism, which has never reported in such CTTBG
moir\'{e} systems.

\begin{acknowledgments}

The work at PKU was supported by the National Key Research and
Development Program of China (Grant No. 2021YFA1401900, 2019YFA0308402
and 2019YFA0308403), the Innovation Program for Quantum Science and
Technology (Grant No. 2021ZD0302602 and 2021ZD0302403), and the
National Natural Science Foundation of China (Grant No. 11934001,
92265106 and 11921005). J.-H.C acknowledges technical support form
Peking Nanofab.

\end{acknowledgments}

\bibliography{./bib_Graphene}

\end{document}